# Designing Magnetic Topological Insulator Trilayers for Highly-Efficient Spin-Orbit Torque Switching

Ling-Jie Zhou[1], Deyi Zhuo[1], Han Tay[1], Zi-Jie Yan[1], Pu Xiao[1], Xiaoda Liu[1], Bomin Zhang[1], and Cui-Zu Chang[1]

[1] Department of Physics, The Pennsylvania State University, University Park, PA 16802, USA

Corresponding authors: cxc955@psu.edu (C.-Z. C.).

**Abstract: Spin-orbit torque (SOT) enables efficient electrical control of magnetization, offering a pathway towards low-power spintronic devices. Magnetic topological insulators (TIs), with spin-momentum-locked surface states and intrinsic ferromagnetism, provide a unique platform for realizing SOT switching of edge current chirality in quantum anomalous Hall (QAH) insulators. In this work, we employ molecular beam epitaxy to synthesize a series of magnetic TI trilayers with controlled layer thicknesses on heat-treated $SrTiO_3$(111) substrates. Electrical transport measurements reveal that SOT-driven magnetization reversal and the associated switching of edge current chirality are governed by the $SrTiO_3$(111) substrate-induced charging effect, which generates an asymmetric chemical-potential alignment between the top and bottom magnetic TI layers. Furthermore, we demonstrate that the switching polarity and efficiency can be tuned through heterostructure design, gate voltage, and in-plane magnetic field, consistent with SOT symmetry. These findings identify chemical potential asymmetry as the origin of the large SOT switching ratio in magnetic TI trilayers and establish a route for electrical control of edge current chirality in QAH insulators. This work advances the understanding of SOT switching mechanism in magnetic topological materials and paves the way for next-generation, energy-efficient QAH-based logic and memory devices.**

**Main text:** The electrical control of magnetic states has long been a central goal in condensed matter physics and spintronics [1-6]. Magnetic materials are fundamental to spin-based information technologies, such as magnetic random-access memory (MRAM), in which data bits are stored by aligning two ferromagnetic layers separated by a tunneling barrier [7]. Spin-transfer torque (STT) has been used to switch magnetization by injecting spin-polarized currents through an insulating barrier [8-11], but the high current densities required for STT can limit device endurance. However, spin-orbit torque (SOT) offers an alternative route by using spin-orbit coupling (SOC) in adjacent nonmagnetic layers to generate spin accumulation via the spin Hall or Rashba-Edelstein effects [1-3, 12-14]. This mechanism enables magnetization switching without passing current through the insulating barrier, greatly enhancing energy efficiency and device reliability [1, 15]. Therefore, SOT has become a cornerstone for developing next-generation nonvolatile memories and logic devices, stimulating intensive exploration of new materials that can serve as highly efficient spin sources.

A wide range of materials has been explored for SOT-based devices, including heavy metals [13, 14, 16], topological insulators (TIs) [17-21], and transition-metal dichalcogenides [22, 23]. Among them, TIs have emerged as particularly promising because their Dirac surface states possess spin-momentum locking, which intrinsically links charge and spin currents [24-26]. When doped with transition metal elements such as Cr and/or V, TIs exhibit long-range ferromagnetism with perpendicular magnetic anisotropy [27-31], an essential feature for scalable MRAM architectures. The breaking of time-reversal symmetry in these magnetic TIs opens a magnetic exchange gap at the Dirac point, giving rise to the quantum anomalous Hall (QAH) insulators with dissipation-free chiral edge channels [28-30, 32-34]. The interplay between magnetism and chiral edge currents enables electrical control of edge chirality through SOT-induced magnetization reversal. Our recent experiments demonstrated current-driven switching of edge current chirality in magnetic TI trilayers with the QAH state [6], marking a major step towards realizing electrically controllable QAH devices. However, the microscopic mechanisms underlying SOT-driven switching of edge current chirality in QAH insulators remain elusive. In magnetic TI trilayers, current injection typically induces opposite spin accumulations at the top and bottom surfaces, canceling the total SOT and yielding a vanishing switching ratio. In contrast, our experimental results reveal substantial magnetic domain reversal, leading to SOT switching of the edge current chirality in QAH insulators [6].

In this work, we systematically investigate the microscopic origin of SOT-driven magnetization switching in magnetic TI trilayers. Using molecular beam epitaxy (MBE), we fabricate a series of high-quality magnetic TI trilayers with precisely controlled layer thicknesses. Through electrical transport measurements, we reveal that magnetization reversal and the associated edge current chirality switching are governed by a substrate-induced charging effect. This effect, arising from the $SrTiO_3$(111) substrate [27-29, 35], creates an asymmetric chemical-potential alignment between the top and bottom magnetic TI layers, leading to unequal spin accumulation and markedly different SOT efficiencies. We further demonstrate that the switching polarity and magnitude can be tuned via heterostructure design, gate voltage, and in-plane magnetic field, consistent with SOT symmetry. These findings identify chemical potential asymmetry as the key mechanism underlying highly-efficient SOT switching and provide direct evidence for the controllable coupling between ferromagnetism and topological surface states in QAH insulators.

All magnetic TI trilayers used in this work are grown on heat-treated $SrTiO_3$(111) substrates in a commercial MBE chamber (Omicron Lab 10) with a base pressure below $2 \times 10^{-10}$ mbar. To examine the influence of the substrate, control trilayer samples are also grown on InP(111)A substrates. Each trilayer sample consists of $m$ quintuple layer (QL) Cr-doped $(Bi,Sb)_2Te_3$ /4 QL $(Bi,Sb)_2Te_3$/$n$ QL Cr-doped $(Bi,Sb)_2Te_3$, denoted as the ($m$4$n$) heterostructure (Fig. 1a). The Cr dopants induce strong out-of-plane ferromagnetism while simultaneously decreasing the SOC. With heavy Cr doping, the Cr-doped $(Bi,Sb)_2Te_3$ layer is a trivial ferromagnetic insulator [36-38], so the middle undoped $(Bi,Sb)_2Te_3$ spacer layer hosts topological surface states across the Cr-doped $(Bi,Sb)_2Te_3$ /$(Bi,Sb)_2Te_3$ interfaces. The ferromagnetic order in Cr-doped $(Bi,Sb)_2Te_3$ layers breaks time-reversal symmetry and opens a magnetic exchange gap at the Dirac point, giving rise to the chiral edge channel and the QAH effect (Fig. 1a) [28-30, 32-34]. The Bi/Sb ratio is optimized to set the chemical potential of the entire trilayer near the charge-neutral point, and a back gate is used to tune the carrier density finely. When either $m$ or $n$ is zero, the trilayer structure reduces to a bilayer, in which one surface hosts a gapless Dirac cone, whereas the other hosts gapped Dirac surface states (Fig. 1b,c). All Hall bar devices with a width $w$ of ~2 μm and an aspect ratio $l/w$ of ~4 are fabricated using a two-step electron-beam lithography and argon-ion milling [6, 39]. Electrical transport measurements are performed in a Physical Property Measurement System (Quantum Design DynaCool, 1.7 K, 9 T). More details on MBE growth, device fabrication, and electrical transport measurements are provided in the Supporting Information.

To investigate the SOT switching mechanism in magnetic TI trilayers, we employ MBE to fabricate a series of magnetic TI samples with varied layer thicknesses, specifically the (043), (343), (342), and (340) heterostructures. We first perform magnetotransport measurements to characterize the sample quality at $T = 2$ K (Fig. 1d-g). At the charge-neutral point, i.e., $V_g = V_g^0$, the zero-magnetic-field Hall resistance $\rho_{yx}(0)$ is ~2.972 kΩ, ~21.148 kΩ, ~24.846 kΩ, and ~3.639 kΩ for the (043), (343), (342), and (340) heterostructures, respectively. The corresponding zero-magnetic-field longitudinal resistance $\rho_{xx}(0)$ are ~14.521 kΩ, ~8.690 kΩ, ~5.824 kΩ, and ~13.384 kΩ for the (043), (343), (342), and (340) heterostructures. The two trilayer samples, i.e., the (343) and (342) heterostructures, exhibit nearly quantized Hall resistance, with $\rho_{yx}(0)$ reaching 0.819 $h/e^2$ and 0.963 $h/e^2$ and corresponding Hall angles α of ~67.6° and ~76.8°, respectively, confirming the appearance of the QAH state at $T = 2$ K in these two trilayer samples [30].

In contrast, the two bilayer samples, i.e., the (043) and (340) heterostructures, exhibit much smaller $\rho_{yx}(0)$ of ~0.115 $h/e^2$ and ~0.141 $h/e^2$, with corresponding Hall angles α of ~11.3° and ~15.1°, respectively. The significantly reduced $\rho_{yx}(0)$ and Hall angle α are attributed to the presence of gapless Dirac surface states on one of the surfaces in the bilayer samples (Fig. 1b, c) [40]. We note that for the (342) trilayer, a subtle kink is observed near the coercive field (Fig. 1f), presumably due to reduced perpendicular magnetic anisotropy caused by the thinner bottom Cr-doped $(Bi,Sb)_2Te_3$ layer. The appearance of the QAH state in the (343) and (342) trilayers is further confirmed by the ($V_g$ - $V_g^0$) dependent $\rho_{yx}(0)$ and $\rho_{xx}(0)$ behaviors. Specifically, $\rho_{yx}(0)$ exhibits a peak approaching $h/e^2$. At the same time, $\rho_{xx}(0)$ shows a dip near $V_g = V_g^0$ (Fig.1i, j). However, the (043) and (340) bilayers exhibit pronounced peaks in both $\rho_{yx}(0)$ and $\rho_{xx}(0)$ near $V_g = V_g^0$, indicating the absence of the QAH state in these two bilayer samples (Fig. 1h, k).

Next, we examine the SOT-induced magnetization switching in these four heterostructures. Figure 2a shows a schematic of the SOT-induced magnetization switching in a magnetic TI trilayer. An in-plane magnetic field $\mu_0 H_x$ is applied along the current direction. As noted above, the top and bottom heavily Cr-doped $(Bi,Sb)_2Te_3$ layers are trivial ferromagnetic insulators, leading to the formation of topological surface states at the top and bottom Cr-doped $(Bi,Sb)_2Te_3/(Bi,Sb)_2Te_3$ interfaces [30, 36, 37]. By injecting a current pulse along the $x$-axis, the Rashba-Edelstein effect induced by the helical Dirac surface states leads to the accumulation of spin $\boldsymbol{S}_y$ along the $y$-axis at both the top and bottom Cr-doped $(Bi,Sb)_2Te_3/(Bi,Sb)_2Te_3$ interfaces [1, 17, 20, 41]. This spin accumulation exerts both a damping-like torque $\boldsymbol{\tau}_{\mathbf{DL}} \propto \boldsymbol{M}\times(\boldsymbol{M}\times\boldsymbol{S}_y) = \boldsymbol{M}\times\mu_0\boldsymbol{H}_{\mathbf{DL}}$ and a field-like torque $\boldsymbol{\tau}_{\mathbf{FL}} \propto \boldsymbol{M}\times\boldsymbol{S}_y = \boldsymbol{M}\times\mu_0\boldsymbol{H}_{\mathbf{FL}}$. Here, $\mu_0\boldsymbol{H}_{\mathbf{DL}}$ and $\mu_0\boldsymbol{H}_{\mathbf{FL}}$ denote the effective magnetic fields generated by the damping-like and field-like torques, respectively, with $\mu_0\boldsymbol{H}_{\mathbf{DL}} \propto \boldsymbol{M}\times\boldsymbol{S}_y$ and $\mu_0\boldsymbol{H}_{\mathbf{FL}} \propto \boldsymbol{S}_y$. $\mu_0\boldsymbol{H}_{\mathbf{DL}}$ is oriented along the $x$-axis, and changes sign upon magnetization reversal, whereas $\mu_0\boldsymbol{H}_{\mathbf{FL}}$ is oriented along the $y$-axis and remains even with respect to $\boldsymbol{M}$. Therefore, to achieve SOT-induced magnetization switching, an external magnetic field along the $x$-axis, $\mu_0H_x$, is required [1, 13]. Moreover, the opposite helicities of the topological surface states at the top and bottom Cr-doped $(Bi,Sb)_2Te_3/(Bi,Sb)_2Te_3$ interfaces generate opposite spin accumulations, resulting in reversed torques and opposite switching directions.

In our current-pulse-induced SOT switching measurements, a direct current (DC) pulse and an alternating current (AC) are applied in parallel to the source contact of the Hall bar device (Fig. 2b). The DC pulse $I_x$ with a duration of ~5 milliseconds is applied along the $x$-axis of the Hall bar device. After injecting this pulse, the device is allowed to balance for 20 seconds before the Hall resistance $\rho_{yx}$ is measured using a standard lock-in technique with a 1μA AC. The Hall bar devices become charged and stabilized after the first current pulse due to the property of the $SrTiO_3(111)$ substrate. The charging effect, induced by the DC pulse, originates from the $SrTiO_3(111)$ substrate (Fig. S1) and is also responsible for the hysteresis in the $V_g$-dependent transport measurements [42].

Figure 2c-f shows the dependence of the normalized Hall resistance $\rho_{yx}/\rho_{yx}(0)$ on the DC pulse $I_x$ for the (043), (343), (342), and (340) heterostructures, respectively. The values of $\rho_{yx}(0)$ are measured after applying the DC pulse $I_x$ at $(V_g - V_g^0) = -20$ V and under an optimal in-plane

magnetic field $\mu_0 H_x^*$. Here, $\mu_0 H_x^*$ is defined as the in-plane magnetic field where the switching ratio $\Delta\rho_{yx}/\rho_{yx}(0)$ reaches its maximum (Fig. 3). $\Delta\rho_{yx}$ characterizes the maximum switched Hall resistance and is obtained by averaging the $[\rho_{yx}(+I_x)-\rho_{yx}(-I_x)]/2$ over DC pulses $I_x > 150$ μA, where the switched Hall resistance is saturated, and the switching ratio $\Delta\rho_{yx}/\rho_{yx}(0)$ measures the fraction of the Hall resistance switched by the DC pulse. $\rho_{yx}(0)$ are ~266.5 Ω, ~912.9 Ω, ~2.645 kΩ, ~215.7 Ω for the (043), (343), (342), and (340) heterostructures, respectively. We find that the switching ratios $\Delta\rho_{yx}/\rho_{yx}(0)$ are ~70.1% and ~68.2% for the (043) and (343) heterostructures, respectively, whereas they are significantly smaller, specifically ~16.0% and ~28.9% for the (342) and (340) heterostructures, respectively. Moreover, the observed reversal of the switching polarity is consistent with the SOT mechanism discussed above.

Next, we summarize our key observations. First, the switching ratio of the (043) heterostructure is significantly larger than that of the (340) heterostructure, despite their purely inverted configuration. Second, the (343) heterostructure exhibits a similarly high switching ratio as the (043) heterostructure, despite the top magnetic TI layer being expected to counteract the switching direction. When the bottom magnetic TI layer thickness is reduced to 2 QL in the (342) heterostructure, the switching behavior changes abruptly, with both the switching polarity and ratio resembling those of the (340) heterostructure. We attribute this observation to ferromagnetic interlayer coupling between the top and bottom magnetic TI layers. The bottom 3 QL Cr-doped $(Bi,Sb)_2Te_3$ layer exhibits stronger switching than the top 3 QL Cr-doped $(Bi,Sb)_2Te_3$ layer. Through interlayer exchange coupling, the bottom 3 QL Cr-doped $(Bi,Sb)_2Te_3$ layer thus governs the switching ratio of the (343) heterostructure. When the thickness of the bottom Cr-doped $(Bi,Sb)_2Te_3$ layer is reduced to 2 QL, the weakened perpendicular magnetic anisotropy allows the top 3 QL Cr-doped $(Bi,Sb)_2Te_3$ layer to dominate the magnetic ordering, resulting in comparable switching ratios in the (342) and (340) heterostructures.

To further confirm that SOT drives the observed electrical switching, we measure the $I_x$-dependent $\rho_{yx}/\rho_{yx}(0)$ curves for the (043), (343), (342), and (340) heterostructures under different $\mu_0H_x$ (Fig. 3a-d). At $\mu_0H_x$ = 0 T, none of the four heterostructures exhibits electrical switching. For the (043) and (343) heterostructures, both $\rho_{yx}$ and $\boldsymbol{M}$ switch to the $+z$ ($-z$) axis when $I_x$ and $\mu_0H_x$ are parallel (antiparallel) (Fig. 3a,b). In contrast, the electrical switching polarity is reversed in the (342) and (340) heterostructures (Fig. 3c,d). We summarize the switching ratio $\Delta\rho_{yx}/\rho_{yx}(0)$ as a function of $\mu_0H_x$ for the (043), (343), (342), and (340) heterostructures (Fig. 3e-h). For all four heterostructures, $|\Delta\rho_{yx}/\rho_{yx}(0)|$ first increases with increasing $\mu_0H_x$ and then decreases. The optimal in-plane magnetic field for SOT switching $\mu_0H_x^*$ is ~0.09 T, ~0.06 T, ~0.03 T, and ~0.05 T for the (043), (343), (342), and (340) heterostructures, respectively (Fig. 3e-h). As noted above, an in-plane magnetic field $\mu_0H_x$ is required to create an energy difference between the $\pm\boldsymbol{M}$ states and is crucial for SOT switching. A moderate $\mu_0H_x$ enhances the SOT switching ratio, whereas a large $\mu_0H_x$ tilts the magnetization into the sample plane, thereby reducing the SOT switching efficiency.

Finally, we discuss the physical origin of SOT switching in magnetic TI trilayers, focusing on the interplay between the switching behaviors of the bottom and top Cr-doped $(Bi,Sb)_2Te_3$ layers. By applying DC pulse injection during SOT switching, the magnetic TI trilayers are driven into the hole-doped regime due to the charging effect induced by the $SrTiO_3(111)$ substrate (Figs. S1 to S3). Because the bottom Cr-doped $(Bi,Sb)_2Te_3$ layer is in direct contact with the $SrTiO_3(111)$ substrate, it experiences the strongest charging effect, which shifts its chemical potential $E_F$ further downward towards the bulk valence band (Fig. 4a). In contrast, the top Cr-doped $(Bi,Sb)_2Te_3$ layer is spatially separated from the $SrTiO_3(111)$ substrate by the middle $(Bi,Sb)_2Te_3$ spacer layer and the bottom Cr-doped $(Bi,Sb)_2Te_3$ layer, so it is partially screened from the substrate-induced electric field. The asymmetry in $E_F$ between the top and bottom Cr-doped $(Bi,Sb)_2Te_3$ layers results in unequal spin accumulation and asymmetric SOT switching ratios (Fig. 4a). Because the charging effect during DC pulse injection is difficult to characterize directly, we investigate the $E_F$ asymmetry through electrostatic gating and transport measurements. The asymmetric chemical potential is further corroborated by the $\mu_0 H_\perp$-dependent $\rho_{yx}$ curves of the (043) and (340) heterostructures at $(V_g - V_g^0) = 0$ V and ± 100 V (Fig. 4b,c). For $(V_g - V_g^0) = -100$ V, i.e., the *p*-doped regime, the ferromagnetism is enhanced, whereas for $(V_g - V_g^0) = +100$ V, i.e., the *n*-doped regime, the ferromagnetism is weakened [43]. Figure 4d,e shows the $(V_g - V_g^0)$ dependence of the coercive field $\mu_0 H_c$. Although both the (043) and (340) heterostructures exhibit similar trends, $\mu_0 H_c$ saturates at ~0.086 T for the (043) heterostructure, but only ~0.055 T for the (340) heterostructure. Because the Hall hysteresis loops at $(V_g - V_g^0) = 0$ V are nearly identical, the observed asymmetry cannot be attributed to MBE growth-induced variations reported in a prior study of $(Bi,Sb)_2Te_3$ films on InP(111) substrates [44]. These results indicate that the bottom Cr-doped $(Bi,Sb)_2Te_3$ layer is more strongly charged than the top Cr-doped $(Bi,Sb)_2Te_3$ layer, resulting in larger spin accumulation and an enhanced SOT switching ratio.

To further examine how the $E_F$ asymmetry affects the SOT switching ratio, we perform current-pulse-induced SOT switching measurements on the (043), (343), (342), and (340) heterostructures at different $E_F$ (Fig. 4f-i). A smaller $\rho_{yx}(0)$ value corresponds to stronger hole doping, indicating a lower $E_F$. We note that all four heterostructures consistently charge towards the hole-doped side after the first DC pulse and remain stable thereafter, even when initially tuned to the charge-neutral point or electron-doped regime. This current pulse-induced charging effect is absent in our heterostructures grown on InP(111)A substrates, suggesting that it originates from the complex dynamic response of the $SrTiO_3(111)$ substrate (Fig. S3).

With increasing hole doping, the switching ratio increases from ~60.7% to ~72.7% for the (043) heterostructure and ~60.6% to ~66.0% for the (343) heterostructure (Fig. 4f,g). In contrast, the (342) and (340) heterostructures, which exhibit opposite switching polarity, show reduced SOT switching ratios, from 15.3% to 5.4% for the (342) heterostructure and from 24.9% to 20.6% for the (340) heterostructure (Fig. 4h, i). These observations can be understood as follows: when $E_F$ moves away from the charge-neutral point, i.e., $(V_g - V_g^0) = 0$ V, more holes accumulate on the bottom Cr-doped $(Bi,Sb)_2Te_3$ layer than on the top, thereby strengthening the SOT switching from the bottom Cr-doped $(Bi,Sb)_2Te_3$ layer. For the (043) and (343) heterostructures, the overall SOT is enhanced, leading to an increase in the SOT switching ratio. In contrast, for the (342) heterostructure, the enhanced SOT from the bottom Cr-doped $(Bi,Sb)_2Te_3$ layer counteracts that from the top Cr-doped $(Bi,Sb)_2Te_3$ layer, leading to a reduced SOT switching ratio. For the (340) heterostructure, the more substantial hole accumulation at the bottom Cr-doped $(Bi,Sb)_2Te_3$ layer directs more current through the bottom gapless surface states, thereby diminishing the SOT switching efficiency of the top Cr-doped $(Bi,Sb)_2Te_3$ layer (Fig. 1c). Therefore, our results demonstrate that the $SrTiO_3(111)$ substrate-induced asymmetry in $E_F$ significantly enhances the SOT switching ratio in the bottom Cr-doped $(Bi,Sb)_2Te_3$ layer, enabling electrical switching of edge current chirality in QAH trilayers (Fig. S2) [6].

To summarize, we use electron-beam lithography to fabricate a series of magnetic TI trilayer Hall bar devices and demonstrate that SOT is an efficient means to manipulate both magnetization and edge current chirality in QAH insulators. We find that the large SOT switching ratio observed in QAH trilayers originates from the $SrTiO_3(111)$ substrate-induced $E_F$ asymmetry, establishing the foundation for electrical control of edge current chirality in QAH insulators. These findings advance our understanding of SOT-driven magnetization switching in magnetic TI multilayers and pave the way for the development of next-generation, energy-efficient QAH-based electronic and spintronic devices.

**Supporting Information.** The Supporting Information is available free of charge on the ACS Publications website.

MBE growth of magnetic TI trilayers, electron-beam lithography of the magnetic TI Hall bar devices, electrical transport measurements, more discussion on the $SrTiO_3$ substrate-induced charging effect and electrical switching of edge-current chirality in QAH insulators, and more transport results of magnetic TI trilayers.

**Author contributions:** CZC conceived and supervised the experiments. DZ, ZJY, HT, and PX grew all magnetic TI samples. LJZ, XL, and BZ fabricated all Hall bar devices and performed PPMS measurements. LJZ and CZC analyzed the data and wrote the manuscript with input from all authors.

**Notes:** The authors declare no competing financial interest.

**Acknowledgments:** We thank C. Liu, F. Xue, and K. Yang for the helpful discussions. This work is primarily supported by the ONR award (N000142412133), including MBE growth, device fabrication, and sample characterization. The PPMS measurements are partially supported by the NSF grant (DMR-2241327) and the ARO grant (W911NF2210159). C. -Z. C. acknowledges support from the Gordon and Betty Moore Foundation's EPiQS Initiative (GBMF9063 to C. -Z. C.).

**Data availability:** The data that support the findings of this article are openly available at Zenodo (https://doi.org/10.5281/zenodo.19478382).

**Figures and figure captions:**

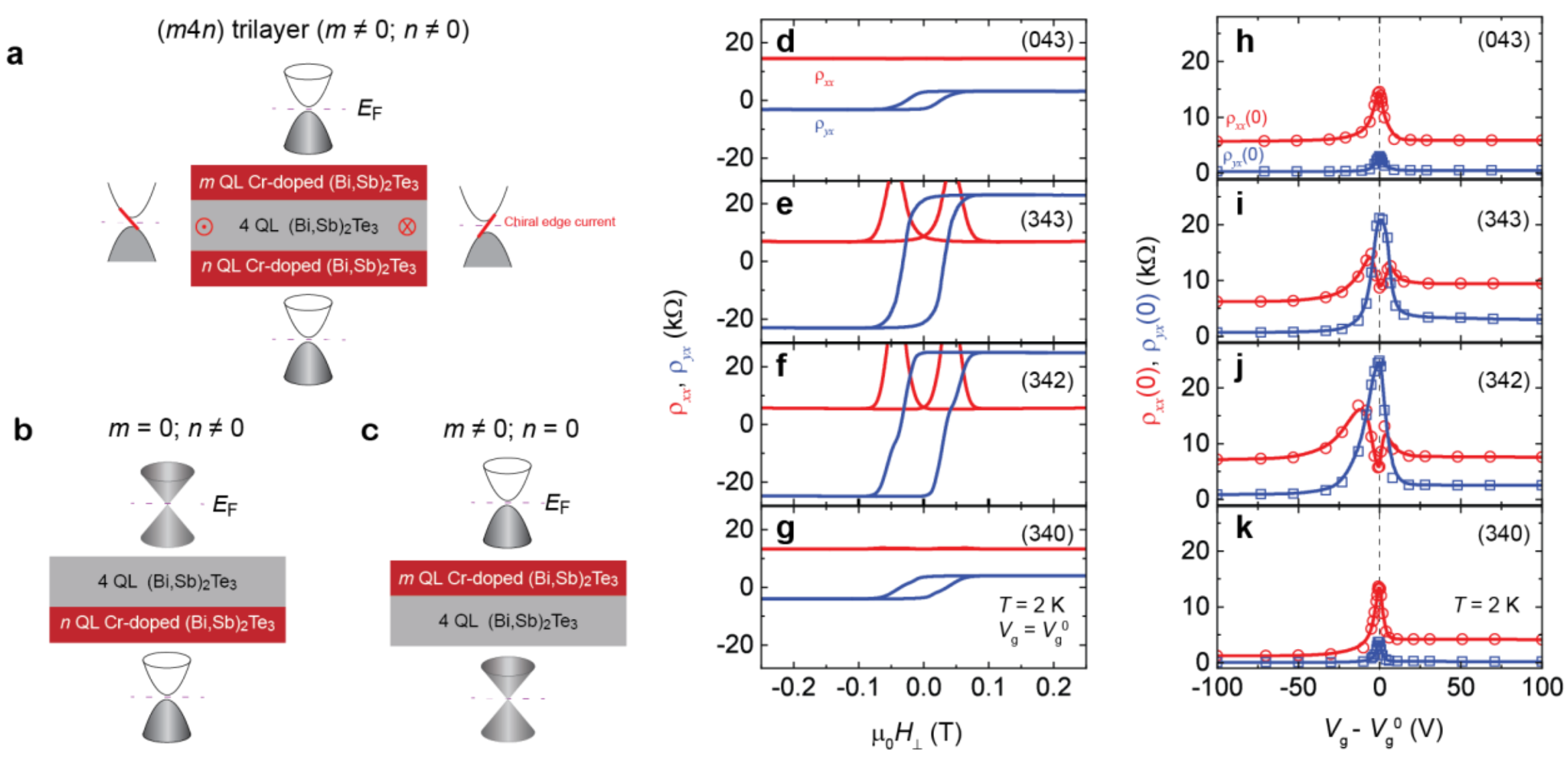

**Fig. 1| Transport results of magnetic TI (*m*4*n*) trilayers. a,** Schematic of a (*m*4*n*) trilayer structure, consisting of *m* QL Cr-doped $(Bi,Sb)_2Te_3$ / 4 QL $(Bi,Sb)_2Te_3$ / *n* QL Cr-doped $(Bi,Sb)_2Te_3$. The two Cr-doped $(Bi,Sb)_2Te_3$ layers open magnetic exchange gaps in the top and bottom surface Dirac surface states, while the chiral edge current circulates along the trilayer edges. **b**, Schematic of the bilayer structure for $m = 0$. **c**, Schematic of the bilayer structure for $n = 0$. **d-g**, $\mu_0 H$-dependent $\rho_{xx}$ (red) and $\rho_{yx}$ (blue) of the (043) (**d**), (343) (**e**), (342) (**f**), and (340) (**g**) heterostructures. at $V_g = V_g^0$ and $T = 2$ K. **h-k**, ($V_g$-$V_g^0$)-dependent $\rho_{xx}(0)$ (red circles) and $\rho_{yx}(0)$ (blue squares) of the (043) (**h**), (343) (**i**), (342) (**j**), and (340) (**k**) heterostructures at $\mu_0 H_\perp = 0$ T and $T = 2$ K. The values of $V_g^0$ are +1 V, +5 V, +2 V, and -1 V for the (043), (343), (342), and (340) heterostructures, respectively.

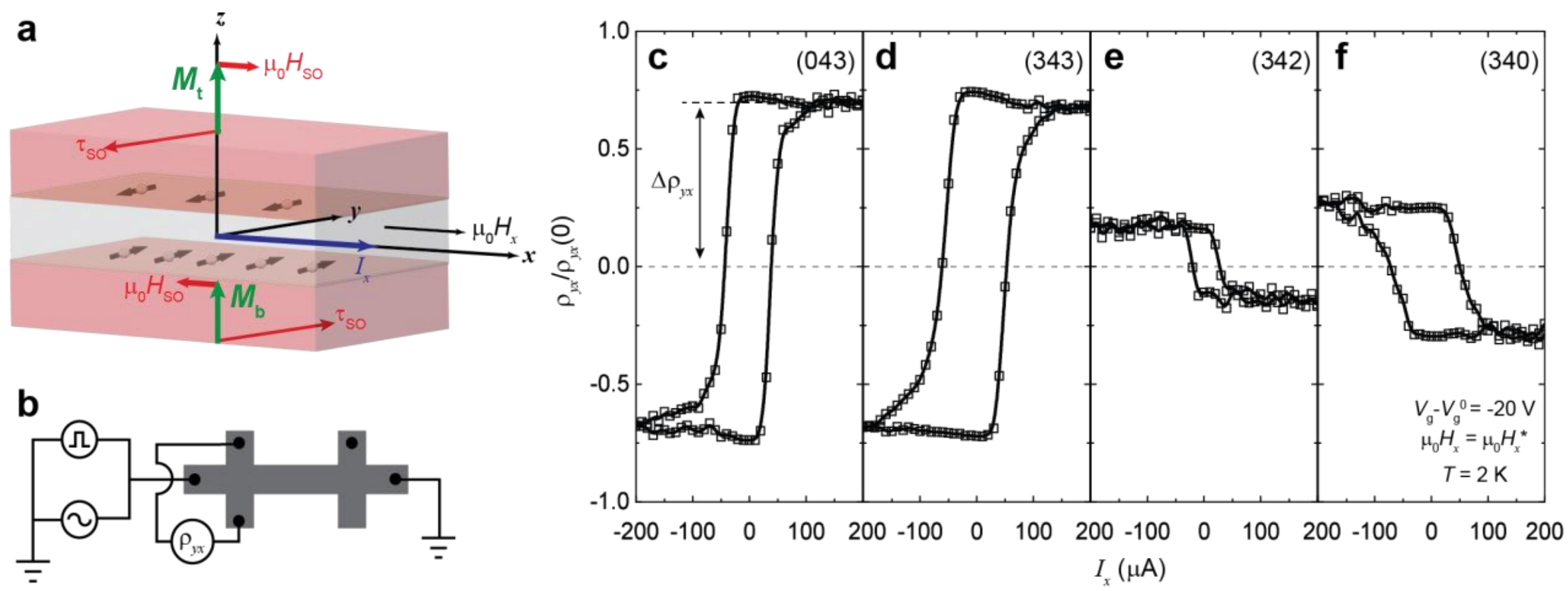

**Fig. 2| SOT switching in magnetic TI (*m*4*n*) trilayers. a**, Schematic of SOT switching in a magnetic TI trilayer induced by a DC pulse $I_x$ under an external in-plane magnetic field $\mu_0 H_x$. **b**, Circuit diagram for the current-pulse-induced SOT switching measurements. A DC pulse $I_x$ with a duration of ~5 milliseconds is applied along the *x*-axis of the Hall bar device. After each pulse, the device is allowed to relax for 20 seconds before $\rho_{yx}$ is measured using a standard AC lock-in technique with a 1 μA AC. **c-f**, $I_x$-dependent $\rho_{yx}/\rho_{yx}(0)$ of the (043) (**c**), (343) (**d**), (342) (**e**), and (340) (**f**) heterostructures at their respective optimal in-plane magnetic field $\mu_0 H_x^*$. The corresponding $\mu_0 H_x^*$ values are ~0.09 T, ~0.06 T, ~0.03 T, and ~0.05 T for the (043), (343), (342), and (340) heterostructures, respectively. The corresponding $\rho_{yx}(0)$ values are ~266.5 Ω, ~912.9 Ω, ~2.645 kΩ, and ~215.7 Ω for the (043), (343), (342), and (340) heterostructures, respectively. Here, $\rho_{yx}(0)$ is measured after application of the DC pulse $I_x$. All measurements are performed at ($V_g$ - $V_g^0$) = -20 V and $T$ = 2 K.

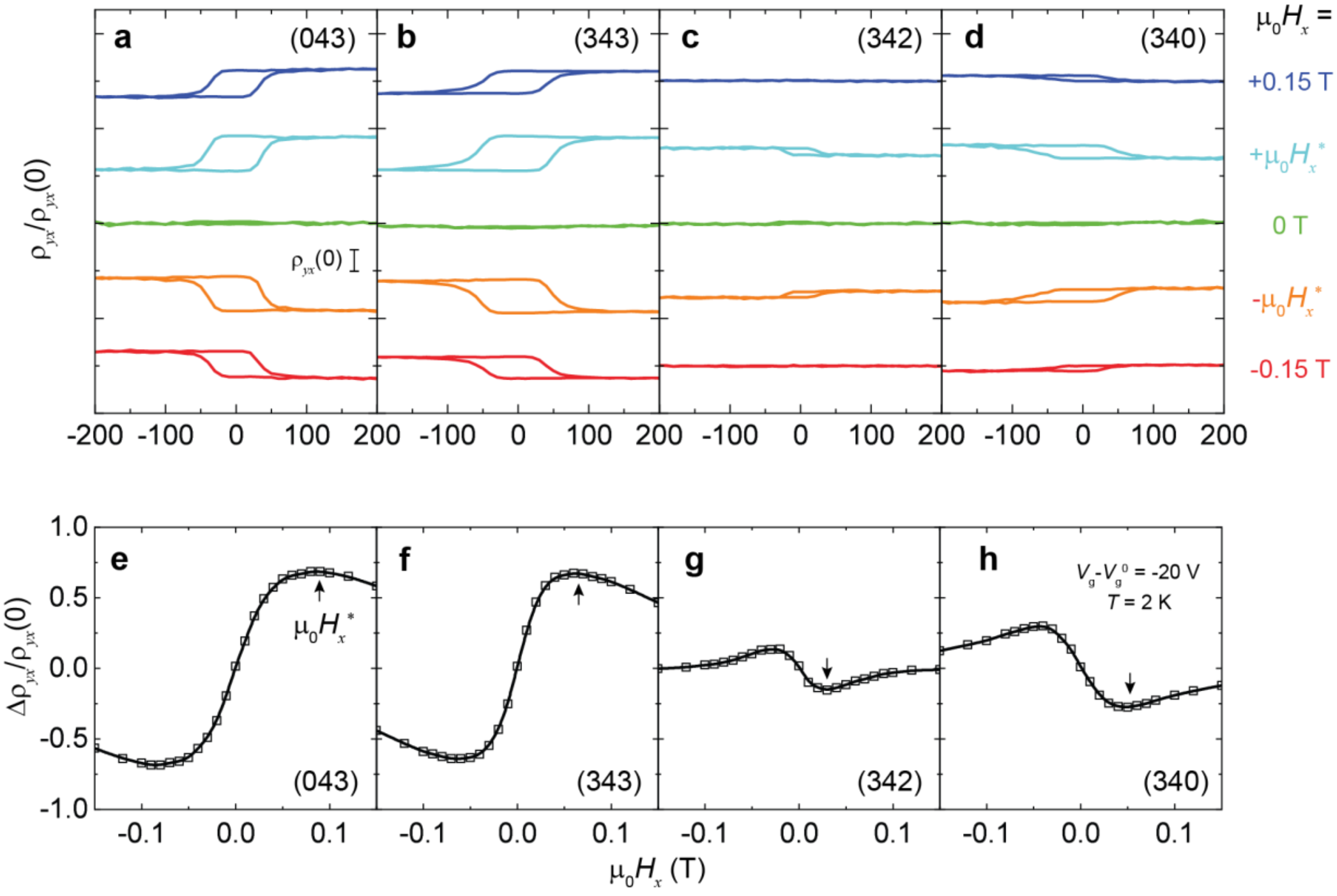

**Fig. 3| DC-pulse-induced SOT switching in magnetic TI (*m*4*n*) trilayers under different $\mu_0 H_x$.** **a-d**, $I_x$-dependent $\rho_{yx}/\rho_{yx}(0)$ for the (043) (**a**), (343) (**b**), (342) (**c**), and (340) (**d**) heterostructures at different $\mu_0 H_x$. **e-h**, $\mu_0 H_x$-dependent $\Delta\rho_{yx}/\rho_{yx}(0)$ for the (043) (**e**), (343) (**f**), (342) (**g**), and (340) (**h**) heterostructures. Black arrows indicate the optimal in-plane magnetic field $\mu_0 H_x^*$. All measurements are performed at ($V_g$ - $V_g^0$) = -20 V and $T$ = 2 K.

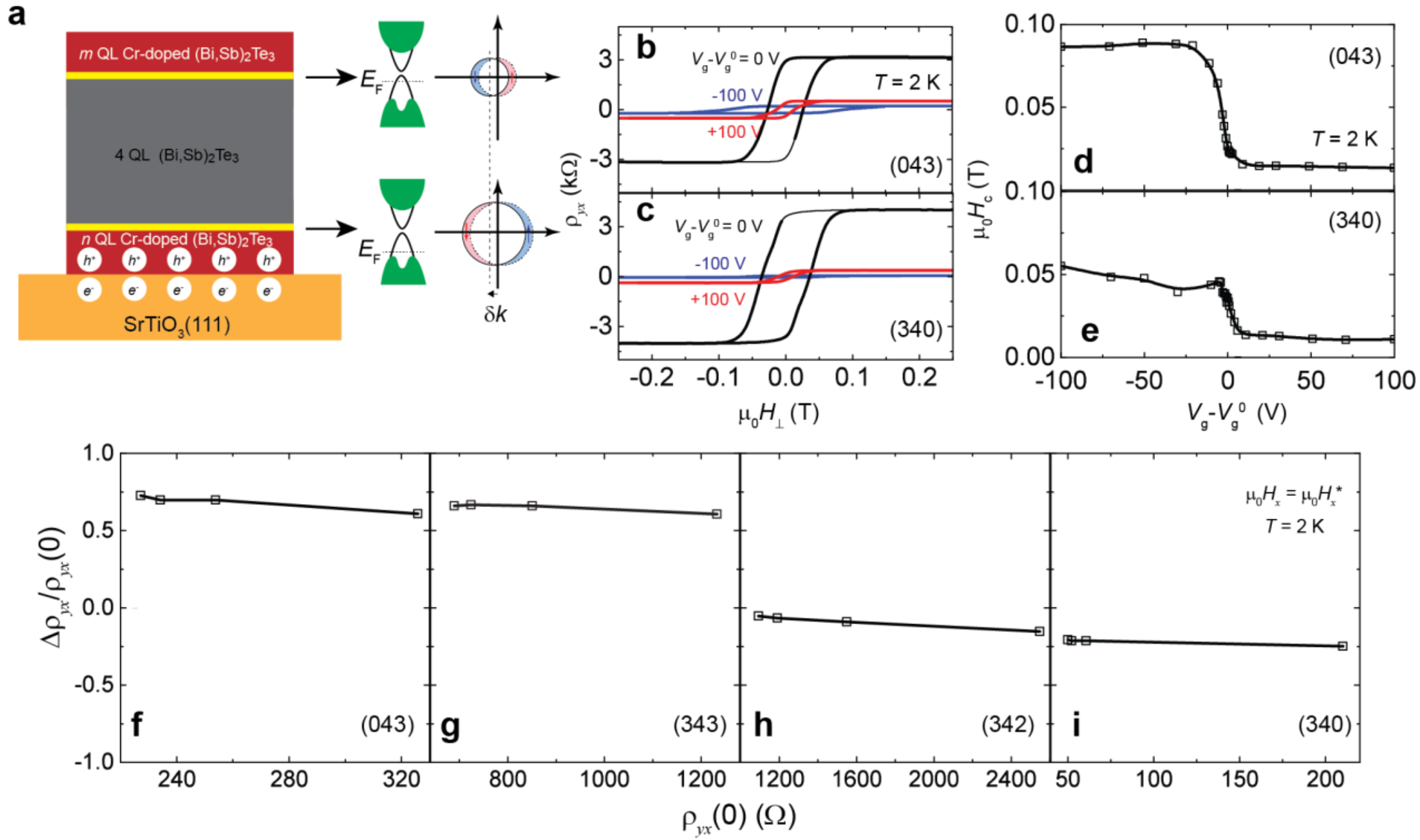

**Fig. 4| Origin of SOT switching in magnetic TI (*m*4*n*) trilayers. a**, Left: Schematic of the charging effect induced by the $SrTiO_3$ substrate. Right: resulting asymmetric chemical potential $E_F$ and spin accumulation during DC pulse injection. The green region is the bulk band, and the parabolic curves are the gapped topological surface states. **b,c**, $\mu_0 H_\perp$-dependent $\rho_{yx}$ measured at ($V_g$ -$V_g^0$) = 0 V and ±100 V for the (043) (**b**) and (340) (**c**) heterostructures. **d, e**, ($V_g$-$V_g^0$) dependence of $\mu_0 H_c$ for the (043) (**d**) and (340) (**e**) heterostructures. **f-i**, $\Delta\rho_{yx}/\rho_{yx}(0)$ plotted as a function of $\rho_{yx}(0)$ for the (043) (**f**), (343) (**g**), (342) (**h**), and (340) (**i**) heterostructures.

## References

1. Manchon, A.; Železný, J.; Miron, I. M.; Jungwirth, T.; Sinova, J.; Thiaville, A.; Garello, K.; Gambardella, P. Current-induced spin-orbit torques in ferromagnetic and antiferromagnetic systems. *Rev. Mod. Phys.* **2019,** 91, 035004.
2. Serlin, M.; Tschirhart, C. L.; Polshyn, H.; Zhang, Y.; Zhu, J.; Watanabe, K.; Taniguchi, T.; Balents, L.; Young, A. F. Intrinsic quantized anomalous Hall effect in a moiré heterostructure. *Science* **2020,** 367, 900-903.
3. Shao, Q.; Li, P.; Liu, L.; Yang, H.; Fukami, S.; Razavi, A.; Wu, H.; Wang, K.; Freimuth, F.; Mokrousov, Y.; Stiles, M. D.; Emori, S.; Hoffmann, A.; Åkerman, J.; Roy, K.; Wang, J. P.; Yang, S. H.; Garello, K.; Zhang, W. Roadmap of Spin–Orbit Torques. *IEEE Trans. Magn.* **2021,** 57, 1-39.
4. Song, C.; Zhang, R.; Liao, L.; Zhou, Y.; Zhou, X.; Chen, R.; You, Y.; Chen, X.; Pan, F. Spin-orbit torques: Materials, mechanisms, performances, and potential applications. *Prog. Mater. Sci.* **2021,** 118, 100761.
5. Tschirhart, C. L.; Redekop, E.; Li, L.; Li, T.; Jiang, S.; Arp, T.; Sheekey, O.; Taniguchi, T.; Watanabe, K.; Huber, M. E.; Mak, K. F.; Shan, J.; Young, A. F. Intrinsic spin Hall torque in a moiré Chern magnet. *Nat. Phys.* **2023,** 19, 807-813.
6. Yuan, W.; Zhou, L.-J.; Yang, K.; Zhao, Y.-F.; Zhang, R.; Yan, Z.; Zhuo, D.; Mei, R.; Wang, Y.; Yi, H.; Chan, M. H. W.; Kayyalha, M.; Liu, C.-X.; Chang, C.-Z. Electrical switching of the edge current chirality in quantum anomalous Hall insulators. *Nat. Mater.* **2024,** 23, 58-64.
7. Evgeny, Y. T.; Oleg, N. M.; Patrick, R. L. Spin-dependent tunnelling in magnetic tunnel junctions. *J. Phys. Condens. Matter.* **2003,** 15, R109.
8. Myers, E. B.; Ralph, D. C.; Katine, J. A.; Louie, R. N.; Buhrman, R. A. Current-Induced Switching of Domains in Magnetic Multilayer Devices. *Science* **1999,** 285, 867-870.
9. Mangin, S.; Ravelosona, D.; Katine, J. A.; Carey, M. J.; Terris, B. D.; Fullerton, E. E. Current-induced magnetization reversal in nanopillars with perpendicular anisotropy. *Nat.*

*Mater.* **2006,** 5, 210-215.

10. Ikeda, S.; Miura, K.; Yamamoto, H.; Mizunuma, K.; Gan, H. D.; Endo, M.; Kanai, S.; Hayakawa, J.; Matsukura, F.; Ohno, H. A perpendicular-anisotropy CoFeB–MgO magnetic tunnel junction. *Nat. Mater.* **2010,** 9, 721-724.
11. Brataas, A.; Kent, A. D.; Ohno, H. Current-induced torques in magnetic materials. *Nat. Mater.* **2012,** 11, 372-381.
12. Liu, L.; Moriyama, T.; Ralph, D. C.; Buhrman, R. A. Spin-Torque Ferromagnetic Resonance Induced by the Spin Hall Effect. *Phys. Rev. Lett.* **2011,** 106, 036601.
13. Liu, L.; Lee, O. J.; Gudmundsen, T. J.; Ralph, D. C.; Buhrman, R. A. Current-Induced Switching of Perpendicularly Magnetized Magnetic Layers Using Spin Torque from the Spin Hall Effect. *Phys. Rev. Lett.* **2012,** 109, 096602.
14. Liu, L.; Pai, C.-F.; Li, Y.; Tseng, H. W.; Ralph, D. C.; Buhrman, R. A. Spin-Torque Switching with the Giant Spin Hall Effect of Tantalum. *Science* **2012,** 336, 555-558.
15. Ji, G.; Zhang, Y.; Chai, Y.; Nan, T. Recent progress on controlling spin-orbit torques by materials design. *npj Spintronics* **2024,** 2, 56.
16. Miron, I. M.; Garello, K.; Gaudin, G.; Zermatten, P.-J.; Costache, M. V.; Auffret, S.; Bandiera, S.; Rodmacq, B.; Schuhl, A.; Gambardella, P. Perpendicular switching of a single ferromagnetic layer induced by in-plane current injection. *Nature* **2011,** 476, 189-193.
17. Han, J.; Richardella, A.; Siddiqui, S. A.; Finley, J.; Samarth, N.; Liu, L. Room-Temperature Spin-Orbit Torque Switching Induced by a Topological Insulator. *Phys. Rev. Lett.* **2017,** 119, 077702.
18. Wu, H.; Zhang, P.; Deng, P.; Lan, Q.; Pan, Q.; Razavi, S. A.; Che, X.; Huang, L.; Dai, B.; Wong, K.; Han, X.; Wang, K. L. Room-Temperature Spin-Orbit Torque from Topological Surface States. *Phys. Rev. Lett.* **2019,** 123, 207205.
19. Mogi, M.; Yasuda, K.; Fujimura, R.; Yoshimi, R.; Ogawa, N.; Tsukazaki, A.; Kawamura, M.; Takahashi, K. S.; Kawasaki, M.; Tokura, Y. Current-induced switching of proximity-induced ferromagnetic surface states in a topological insulator. *Nat. Commun.* **2021,** 12, 1404.

20. Tokura, Y.; Yasuda, K.; Tsukazaki, A. Magnetic topological insulators. *Nat. Rev. Phys.* **2019,** 1, 126-143.

21. Tai, L.; He, H.; Chong, S. K.; Zhang, H.; Huang, H.; Qiu, G.; Ren, Y.; Li, Y.; Yang, H.-Y.; Yang, T.-H.; Dong, X.; Dai, B.; Qu, T.; Shu, Q.; Pan, Q.; Zhang, P.; Xue, F.; Li, J.; Davydov, A. V.; Wang, K. L. Giant Hall Switching by Surface-State-Mediated Spin-Orbit Torque in a Hard Ferromagnetic Topological Insulator. *Adv. Mater.* **2024,** 36, 2406772.

22. MacNeill, D.; Stiehl, G. M.; Guimaraes, M. H. D.; Buhrman, R. A.; Park, J.; Ralph, D. C. Control of spin–orbit torques through crystal symmetry in $WTe_2$/ferromagnet bilayers. *Nat. Phys.* **2017,** 13, 300-305.

23. Kao, I. H.; Muzzio, R.; Zhang, H.; Zhu, M.; Gobbo, J.; Yuan, S.; Weber, D.; Rao, R.; Li, J.; Edgar, J. H.; Goldberger, J. E.; Yan, J.; Mandrus, D. G.; Hwang, J.; Cheng, R.; Katoch, J.; Singh, S. Deterministic switching of a perpendicularly polarized magnet using unconventional spin–orbit torques in $WTe_2$. *Nat. Mater.* **2022,** 21, 1029-1034.

24. Hasan, M. Z.; Kane, C. L. Colloquium: Topological insulators. *Rev. Mod. Phys.* **2010,** 82, 3045-3067.

25. Qi, X.-L.; Zhang, S.-C. Topological insulators and superconductors. *Rev. Mod. Phys.* **2011,** 83, 1057-1110.

26. Mellnik, A. R.; Lee, J. S.; Richardella, A.; Grab, J. L.; Mintun, P. J.; Fischer, M. H.; Vaezi, A.; Manchon, A.; Kim, E. A.; Samarth, N.; Ralph, D. C. Spin-transfer torque generated by a topological insulator. *Nature* **2014,** 511, 449-451.

27. Chang, C.-Z.; Zhang, J.; Liu, M.; Zhang, Z.; Feng, X.; Li, K.; Wang, L.-L.; Chen, X.; Dai, X.; Fang, Z.; Qi, X.-L.; Zhang, S.-C.; Wang, Y.; He, K.; Ma, X.-C.; Xue, Q.-K. Thin Films of Magnetically Doped Topological Insulator with Carrier-Independent Long-Range Ferromagnetic Order. *Adv. Mater.* **2013,** 25, 1065-1070.

28. Chang, C.-Z.; Zhang, J.; Feng, X.; Shen, J.; Zhang, Z.; Guo, M.; Li, K.; Ou, Y.; Wei, P.; Wang, L.-L.; Ji, Z.-Q.; Feng, Y.; Ji, S.; Chen, X.; Jia, J.; Dai, X.; Fang, Z.; Zhang, S.-C.; He, K.; Wang, Y.; Lu, L.; Ma, X.-C.; Xue, Q.-K. Experimental Observation of the Quantum

Anomalous Hall Effect in a Magnetic Topological Insulator. *Science* **2013,** 340, 167-170.

29. Chang, C.-Z.; Zhao, W.; Kim, D. Y.; Zhang, H.; Assaf, B. A.; Heiman, D.; Zhang, S.-C.; Liu, C.; Chan, M. H. W.; Moodera, J. S. High-precision realization of robust quantum anomalous Hall state in a hard ferromagnetic topological insulator. *Nat. Mater.* **2015,** 14, 473-477.
30. Chang, C.-Z.; Liu, C.-X.; MacDonald, A. H. Colloquium: Quantum anomalous Hall effect. *Rev. Mod. Phys.* **2023,** 95, 011002.
31. Ou, Y.; Liu, C.; Jiang, G.; Feng, Y.; Zhao, D.; Wu, W.; Wang, X.-X.; Li, W.; Song, C.; Wang, L.-L.; Wang, W.; Wu, W.; Wang, Y.; He, K.; Ma, X.-C.; Xue, Q.-K. Enhancing the Quantum Anomalous Hall Effect by Magnetic Codoping in a Topological Insulator. *Adv. Mater.* **2018,** 30, 1703062.
32. Checkelsky, J. G.; Yoshimi, R.; Tsukazaki, A.; Takahashi, K. S.; Kozuka, Y.; Falson, J.; Kawasaki, M.; Tokura, Y. Trajectory of the anomalous Hall effect towards the quantized state in a ferromagnetic topological insulator. *Nat. Phys.* **2014,** 10, 731-736.
33. Kou, X.; Guo, S.-T.; Fan, Y.; Pan, L.; Lang, M.; Jiang, Y.; Shao, Q.; Nie, T.; Murata, K.; Tang, J.; Wang, Y.; He, L.; Lee, T.-K.; Lee, W.-L.; Wang, K. L. Scale-Invariant Quantum Anomalous Hall Effect in Magnetic Topological Insulators beyond the Two-Dimensional Limit. *Phys. Rev. Lett.* **2014,** 113, 137201.
34. Mogi, M.; Yoshimi, R.; Tsukazaki, A.; Yasuda, K.; Kozuka, Y.; Takahashi, K. S.; Kawasaki, M.; Tokura, Y. Magnetic modulation doping in topological insulators toward higher-temperature quantum anomalous Hall effect. *Appl. Phys. Lett.* **2015,** 107.
35. Chen, J.; Qin, H. J.; Yang, F.; Liu, J.; Guan, T.; Qu, F. M.; Zhang, G. H.; Shi, J. R.; Xie, X. C.; Yang, C. L.; Wu, K. H.; Li, Y. Q.; Lu, L. Gate-Voltage Control of Chemical Potential and Weak Antilocalization in $Bi_2Se_3$. *Phys. Rev. Lett.* **2010,** 105, 176602.
36. Zhao, Y.-F.; Zhang, R.; Mei, R.; Zhou, L.-J.; Yi, H.; Zhang, Y.-Q.; Yu, J.; Xiao, R.; Wang, K.; Samarth, N.; Chan, M. H. W.; Liu, C.-X.; Chang, C.-Z. Tuning the Chern number in quantum anomalous Hall insulators. *Nature* **2020,** 588, 419-423.
37. Zhao, Y.-F.; Zhang, R.; Zhou, L.-J.; Mei, R.; Yan, Z.-J.; Chan, M. H. W.; Liu, C.-X.; Chang,

C.-Z. Zero Magnetic Field Plateau Phase Transition in Higher Chern Number Quantum Anomalous Hall Insulators. *Phys. Rev. Lett.* **2022,** 128, 216801.

38. Zhou, L.-J.; Zhuo, D.; Mei, R.; Zhao, Y.-F.; Yang, K.; Zhang, R.; Yan, Z.; Tay, H.; Chan, M. H. W.; Liu, C.-X.; Chang, C.-Z. Interlayer coupling induced quantum phase transition in quantum anomalous Hall multilayers. *Phys. Rev. B* **2025,** 111, L201304.
39. Zhou, L.-J.; Mei, R.; Zhao, Y.-F.; Zhang, R.; Zhuo, D.; Yan, Z.-J.; Yuan, W.; Kayyalha, M.; Chan, M. H. W.; Liu, C.-X.; Chang, C.-Z. Confinement-Induced Chiral Edge Channel Interaction in Quantum Anomalous Hall Insulators. *Phys. Rev. Lett.* **2023,** 130, 086201.
40. Mogi, M.; Okamura, Y.; Kawamura, M.; Yoshimi, R.; Yasuda, K.; Tsukazaki, A.; Takahashi, K. S.; Morimoto, T.; Nagaosa, N.; Kawasaki, M.; Takahashi, Y.; Tokura, Y. Experimental signature of the parity anomaly in a semi-magnetic topological insulator. *Nat. Phys.* **2022,** 18, 390-394.
41. Edelstein, V. M. Spin polarization of conduction electrons induced by electric current in two-dimensional asymmetric electron systems. *Solid State Commun.* **1990,** 73, 233-235.
42. Chang, C.-Z.; Zhao, W.; Li, J.; Jain, J. K.; Liu, C.; Moodera, J. S.; Chan, M. H. W. Observation of the Quantum Anomalous Hall Insulator to Anderson Insulator Quantum Phase Transition and its Scaling Behavior. *Phys. Rev. Lett.* **2016,** 117, 126802.
43. Kou, X.; He, L.; Lang, M.; Fan, Y.; Wong, K.; Jiang, Y.; Nie, T.; Jiang, W.; Upadhyaya, P.; Xing, Z.; Wang, Y.; Xiu, F.; Schwartz, R. N.; Wang, K. L. Manipulating Surface-Related Ferromagnetism in Modulation-Doped Topological Insulators. *Nano Lett.* **2013,** 13, 4587-4593.
44. Yoshimi, R.; Yasuda, K.; Tsukazaki, A.; Takahashi, K. S.; Nagaosa, N.; Kawasaki, M.; Tokura, Y. Quantum Hall states stabilized in semi-magnetic bilayers of topological insulators. *Nat. Commun.* **2015,** 6, 8530.